\documentclass[aps,prl,reprint,groupedaddress,nofootinbib]{revtex4-1}
\usepackage[]{natbib}
\usepackage{amsmath, amssymb, amsthm, url}
\usepackage{graphicx}
\usepackage[switch,running,mathlines]{lineno} 
\usepackage{xcolor}  % Allow for color text
\usepackage[normalem]{ulem} % Allow for strikeout text (\sout)
\usepackage{pdfpages} % Allow inclusion of pdf
\usepackage{pgffor} % Allow for loops in control?

\graphicspath{{Figures_condensed/}}

\newcommand{\sgn}{\operatorname{sgn}}

\newcommand{\dadt}{\mathrm{d}a/\mathrm{d}t}
\newcommand{\dadtfrac}{\frac{\mathrm{d}a}{\mathrm{d}t}}

\newcommand{\abar}{\bar{a}}
\newcommand{\aopt}{a_{\textrm{opt}}}
\newcommand{\p}{\partial}
\newcommand{\phiind}{\varphi^{\textrm{(ind)}}}
\newcommand{\phisoc}{\varphi^{\textrm{(soc)}}}
\begin{document}

%\preprint{}

\title{Handicap principle implies emergence of dimorphic ornaments}
%SHORT TITLE \title{Handicap principle explains dimorphic ornaments}



\author{Sara M.~Clifton}
\email[E-mail me at: ]{sclifton@u.northwestern.edu}
\affiliation{Department of Engineering Sciences and Applied Mathematics, Northwestern University, Evanston, Illinois 60208, USA}
\author{Rosemary I.~Braun}
\affiliation{Division of Biostatistics, Northwestern University, Chicago, Illinois 60611, USA}
\affiliation{Northwestern Institute for Complex Systems, Northwestern University, Evanston, Illinois 60208, USA}
\author{Daniel M.~Abrams}
\affiliation{Department of Engineering Sciences and Applied Mathematics, Northwestern University, Evanston, Illinois 60208, USA}
\affiliation{Northwestern Institute for Complex Systems, Northwestern University, Evanston, Illinois 60208, USA}
\affiliation{Department of Physics and Astronomy, Northwestern University, Evanston, Illinois 60208, USA}

%\keywords{ornament | evolution | natural selection | sexual selection | competition | speciation | handicap principle | mathematical model}

\date{\today}

\begin{abstract}
Species spanning the animal kingdom have evolved extravagant and costly ornaments to attract mating partners.  Zahavi's handicap principle offers an elegant explanation for this: ornaments signal individual quality, and must be costly to ensure honest signaling, making mate selection more efficient. Here we incorporate the assumptions of the handicap principle into a mathematical model and show that they are sufficient to explain the heretofore puzzling observation of bimodally distributed ornament sizes in a variety of species.
\end{abstract}

\maketitle

%\linenumbers
\
%%%%%%%%%%%%%%%%%%%%%%%%%%%%%%%%%%%%%%%%%%%%%%
%                                                                                                                                             %
%                                                   BACKGROUND                                                                %
%                                                                                                                                             %
%%%%%%%%%%%%%%%%%%%%%%%%%%%%%%%%%%%%%%%%%%%%%%
\section{Background \label{background}}
Darwin was the first to suggest that both natural and sexual selection play a role in the evolution of mating displays \cite{Darwin:1871}. Natural selection is the shift in population traits based on an individual's ability to survive and gather resources, while sexual selection is the shift in population traits based on an individual's ability to mate with more or better partners. Natural selection alone cannot explain ornaments because they hinder survival and provide little to no benefit to the individual\cite{Andersson:2006jd, CluttonBrock:2007jka,Jones:2009vo}. Darwin hypothesized that female preference for exaggerated mating displays drives the evolution of male ornamentation, but he was unable to explain why females prefer features which clearly handicap the males.

Zahavi's handicap principle attempts to resolve the paradox proposed by Darwin \cite{Zahavi:1975uaa}. It argues that, because costly ornaments hinder survival, only the highest quality individuals can afford significant investment in them. Thus the cost (often correlated with size) of an ornament truthfully advertises the quality of an individual, which makes mate selection easier. There is a large body of evidence that ornaments are indeed costly to the bearer (e.g. \cite{ALLEN:2007it, evans1992aerodynamic, Goyens:2015cm}), that ornaments are honest signals of quality (e.g., \cite{Johnstone:1995wt,blount2003carotenoid}), and that females prefer mates with larger ornaments (e.g. \cite{West:2002cp, Petrie:1994wp, Andersson:1982tr}). 

A variety of theoretical approaches have been used to model the handicap principle \cite{Jones:2009vo, kuijper2012guide, collins1993there, kokko2006unifying, hill2014evolution}. Broad categories include game theoretical approaches (e.g., \cite{gintis2001costly, grafen1990biological}), quantitative genetics (e.g., \cite{iwasa1991evolution, Lande:1981th}), and phenotypic dynamics (e.g., \cite{nowak2006, dieckmann1996dynamical}).  Borrowing and expanding upon ideas from all three methods, we propose a new dynamical systems approach to understanding the evolution of ornaments within a population.  Our model differs from some that search for a single evolutionarily stable strategy (ESS) (e.g., \cite{grafen1990biological}) in that we do not require a unique phenotype for a particular male quality; our method allows for the possibility that an optimal \textit{distribution} of strategies may emerge for a population---even a population of equal quality males.  

\begin{figure}[htb!]
  \centering
    \includegraphics[width=0.6\columnwidth]{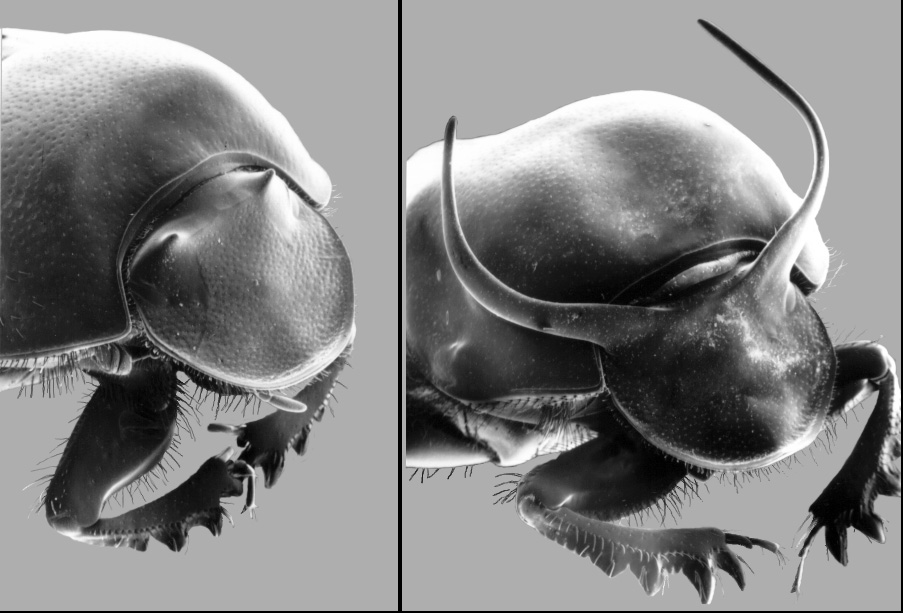}
     \caption{Example of a dimorphic ornament: dung beetles with differing horn lengths (\textit{Onthophagus taurus}, Coleoptera: Scarabaeidae), reprinted from \cite{Emlen:1999vj} with permission.}
    \label{fig:beetlepic}
\end{figure}

Curiously, it has been observed that ornament sizes frequently have bimodal distributions, resulting in distinct small- and large- ``morphs'' in many ornamented species (e.g., \cite{Emlen:1999vj, Aisenberg:2008ir, Tomkins:2005jf}). Figure \ref{fig:beetlepic} illustrates a classic example of ornament dimorphism, the horned dung beetle~\cite{Emlen:1999vj}. While in some cases researchers have identified genetic and environmental factors associated with ornament size variation (e.g., \cite{Glover:2003vl, Glover:2006da}), the splitting into two \textit{distinct} large- and small-ornamented subpopulations (morphs) remains a contentious area of study. 

Some evolutionary theories suggest that variety within the sexes may be due to varied mating strategies such as mimicry, sneaking, or fighting \cite{west1991sexual, gross1996alternative}. However, our model suggests that the handicap principle alone may be sufficient to explain the origin of the observed ornament bimodality.

%%%%%%%%%%%%%%%%%%%%%%%%%%%%%%%%%%%%%%%%%%%%%%
%                                                                                                                                             %
%                                                             MODEL DERIVATION                                             %
%                                                                                                                                             %
%%%%%%%%%%%%%%%%%%%%%%%%%%%%%%%%%%%%%%%%%%%%%% 
\section{Model \label{model}}
With the goal of examining the quantitative implications of the handicap principle, we construct a minimal dynamical systems model for the evolution of extravagant and costly ornaments on animals.  This proposed model incorporates two components of ornament evolution: an intrinsic cost of ornamentation to an individual (natural selection), and a social benefit of relatively large ornaments within a population (sexual selection). We show that on an evolutionary time scale, identically healthy animals can be forced to split into two morphs, one with large ornaments and one with small. 

To express our model, we introduce the idea of a ``reproductive potential'' $\varphi$.  This can be though of as similar to fitness, though our definition differs from the fitness function commonly used in the replicator equation \cite{nowak2006, karev2010mathematical} (we make the relationship between the two explicit in the supplementary information).  Over long time scales the effect of evolution is to select for individuals with higher reproductive potential.

\begin{figure}[htb!]
  \centering
    \includegraphics[width=\columnwidth]{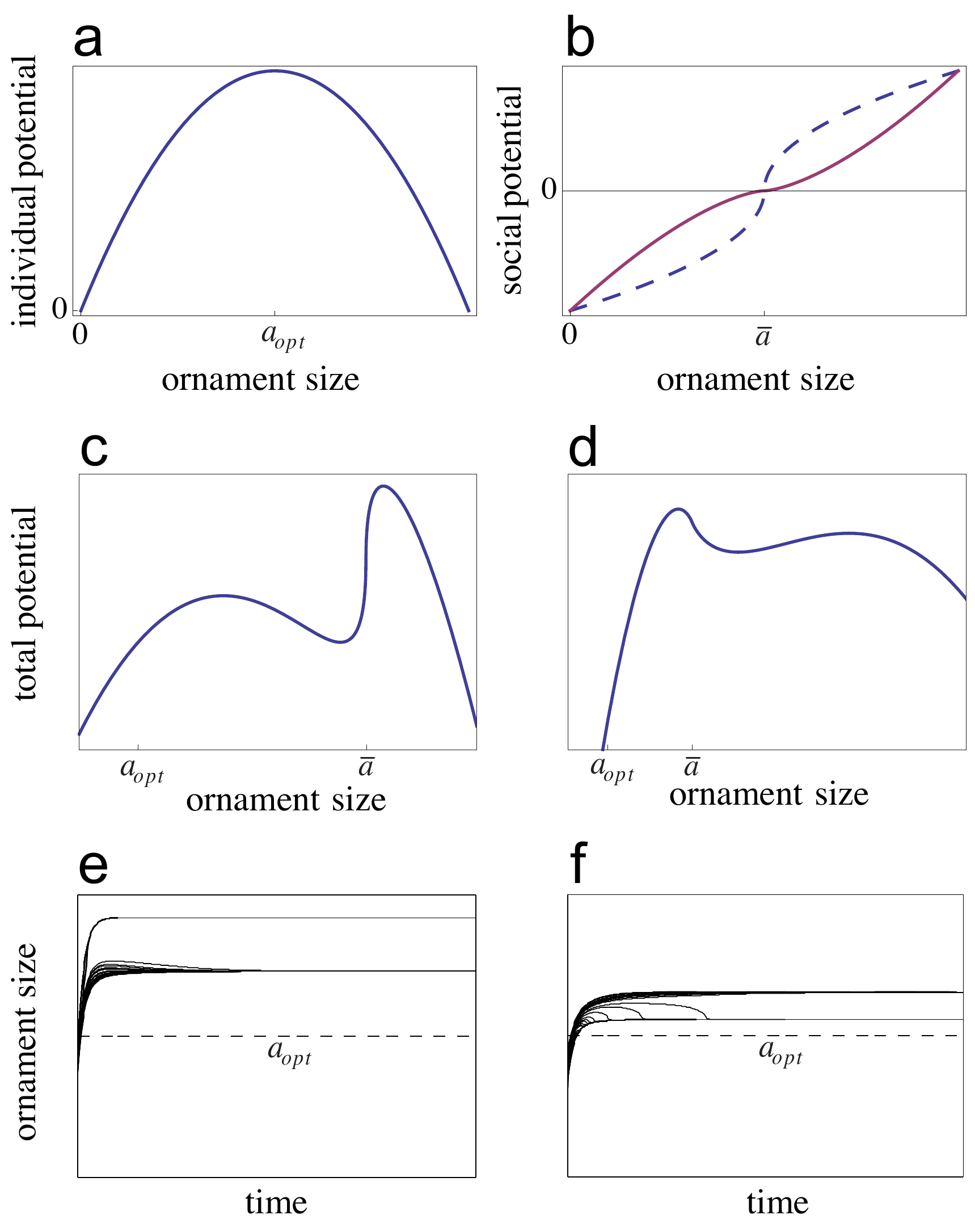}
   % \spacing{1}
    \caption{(Color online) Model derivation and behavior. \textbf{(a)} Example individual potential function, singly peaked at $\aopt$. We use a quadratic function. \textbf{(b)} Example social potential function, antisymmetric about the population mean $\abar$. We use an anti-symmetrized power law such that the shape depends on the social sensitivity $\gamma$ (blue dashed is $\gamma=0.5$; maroon solid is $\gamma=1.5$). \textbf{(c)} Example total reproductive potential function at equilibrium for $\gamma<1$. There are two local maxima corresponding to two distinct morphs, with the larger ornament morph having the highest potential (here $\gamma=0.5$). \textbf{(d)} Example total potential function at equilibrium for $1<\gamma<2$. There are two local maxima corresponding to two distinct morphs, with the smaller ornament morph having the highest potential (here $\gamma=1.5$). Note that the fitness landscape is distinct for each population representative, and representatives are not assumed to be identical. \textbf{(e)} Evolution of $N=100$ population representatives over time for $\gamma=0.5$ and \textbf{(f)} $\gamma=1.5$. The initial conditions were sampled randomly from a normal distribution with mean 0.75 and standard deviation 0.25. The optimal ornament size $\aopt=1.0$, maximum simulation time $t_{max} = 50$, time scaling constant $c=1.0$, and $s=1/2$.}
  %  \spacing{2}
    \label{fig:model}
\end{figure}

Consider an \textit{individual} reproductive potential $\phiind$ of a solitary male with ornament size $a$ (e.g., a deer with ornamental antlers). Some ornaments have practical as well as ornamental value (e.g., anti-predation \cite{Galeotti:2007fo, vandenBrink:2012do}), but have a deleterious effect beyond a certain size. We therefore expect that there exists an optimal ornament size (possibly zero), for which individual potential is maximum, and thus take this to be a singly-peaked function of ornament size.  For simplicity we assume the quadratic form\footnote{This is a generic form for an arbitrary smooth peaked function approximated close to its peak.}
\begin{equation} 
  \phiind = a (2 \aopt- a).
  \label{eq:phiind}
\end{equation}
Following the handicap principle, we expect the optimal ornament size $\aopt=\aopt(h)$ to be an increasing function of ``intrinsic health'' $h$---i.e., healthier individuals can afford larger ornaments. See figure~\ref{fig:model} (a) for the general shape of the individual reproductive potential function.

Next, we consider a \textit{social} reproductive potential $\phisoc$ that captures the effects of competition for partners (i.e., sexual selection). We assume social potential is an increasing function of ornament size
%\footnote{This assumption applies most naturally to inter-sexual selection, ignoring alternative reproductive strategies associated with intra-sexual selection (e.g., cryptic males).} 
because sexual selection often favors larger or more elaborate ornaments \cite{Petrie:1994wp}. For simplicity, and motivated by the ubiquity of power laws in nature \cite{Newman:2005vk, Reed:2002ua}, we choose social potential to be a power of the difference between a male's ornament size and the average herd ornament size. To ensure monotonicity, we force the social reproductive potential to be antisymmetric about the average ornament size. The social potential is then
\begin{equation} 
  \phisoc = \sgn{(a - \abar)}|a - \abar|^{\gamma},
  \label{eq:phisoc}
\end{equation}
where the positive parameter $\gamma$ quantifies the rate at which deviations from the mean influence reproductive potential, $\sgn$ is the sign function, and $\abar$ represents the average ornament size in the population. Loosely speaking, the parameter $\gamma$ tunes female choice; we take this ``female choice" parameter to be effectively constant because female choice may evolve on a slower time scale than male ornamentation \cite{Lande:1981th}. Refer to figure~\ref{fig:model} (b) for an example of the social reproductive potential function. 

Because both natural and sexual selection play a role in the evolution of ornaments \cite{Lande:1981th}, we take total reproductive potential to be the weighted average
\begin{equation} 
  \varphi = s \phisoc + (1-s)  \phiind,
  \label{phitot}
\end{equation}
where $s$ tunes the relative importance of competitive social effects (sexual selection) versus individual effects (natural selection). 
%We show in the supplementary information that a weighted product \cite{getty2006sexually} produces identical qualitative results, so we focus on this case for simplicity of calculations.  
See figure~\ref{fig:model} (c),(d) for examples of total potential functions. 

Assuming that evolutionary forces optimize overall reproductive potential at a rate proportional to the marginal benefit of ornamentation, ornament sizes will follow the dynamics
\begin{equation} 
  \dadtfrac = c \frac{\p \varphi}{\p a} 
  \label{dynsys}
\end{equation}
with time-scaling parameter $c>0$. Note that this model does not presume that individual ornaments explicitly change size: the ``phenotype flux'' $\dadt$  is simply a way of describing how the distribution of ornament sizes in a large animal population changes over long time scales as a result of selection processes.

This results in a piecewise-smooth ordinary differential equation for the ornament size flux,
\begin{equation} 
  \frac{\mathrm{d} a}{\mathrm{d} t} = c \left[ 
    s \gamma \left( 1-\frac{1}{N} \right) 
    \left|a - \abar\right|^{\gamma-1} + 2 (1-s) (\aopt-a)
    \right]. 
\label{eq:finalmodel}
\end{equation}
where $N$ is the population size. Plugging \eqref{eq:finalmodel} into the continuity equation yields a replicator equation for the evolution of the ornament size distribution (see supplementary information).

%To approximate the continuous distribution of ornament size, we consider many instances of \eqref{eq:finalmodel} evolving together, forming a system of $N$ ordinary differential equations. This formulation is more tractable to study than the partial differential replicator equation \eqref{replicator}, yet it yields equivalent results.

%%%%% MODEL FIGURE %%%%%%%%%
%%%%% MODEL FIGURE %%%%%%%%%

%%%%%%%%%%%%%%%%%%%%%%%%%%%%%%%%%%%%%%%%%%%%%%
%                                                                                                                                             %
%                                                       MODEL RESULTS                                                        %
%                                                                                                                                             %
%%%%%%%%%%%%%%%%%%%%%%%%%%%%%%%%%%%%%%%%%%%%%%
\section{Results}
\subsection{Numerical exploration \label{numresults}}
For biologically relevant values of the social sensitivity parameter $\gamma$, our model predicts stratification into distinct phenotypes for a population of identically healthy individuals (i.e., individuals of identical quality). See figure~\ref{fig:model} (e),(f) for the time evolution of ornament size for two representative values of $\gamma$.

For $0 < \gamma < 1$, the ornament sizes stratify into large-ornament and small-ornament groups, with the majority possessing a large-ornament ``morph.''  
%For the special case $\gamma = 1$, equilibrium ornament sizes are uniform.  
For $1 < \gamma < 2$, the population stratifies into large- and small-ornament morphs, but the majority have small ornaments. 
The case $\gamma \ge 2$ is not a reasonable option because we have selected a quadratic form for the local approximation of the individual potential function; any power $\gamma$ exceeding 2 implies sexual selection is the dominant evolutionary force even for extremely large ornaments, an unreasonable assumption. 

These qualitative results are consistent for all $\aopt$ and $0 \le s < 1$. While for clarity we have presented predictions of a specific minimal model, the qualitative results hold for a wide range of models. See the Discussion section for the generality of model predictions.

%%%%%%%%%%%%%%%%%%%%%%%%%%%%%%%%%%%%%%%%%%%%%%
%                                                                                                                                             %
%                                                                 UNIFORM FP                                                     %
%                                                                                                                                             %
%%%%%%%%%%%%%%%%%%%%%%%%%%%%%%%%%%%%%%%%%%%%%%
\subsection{Analytical results \label{anaresults}}
As numerical integration shows that the uniform and two-morph steady states are of interest, we concentrate our analysis on these equilibria. However, it can also be shown graphically that uniform and two-morph steady states are the only possible solutions for a wide range of potential functions (see supplementary information).

\paragraph{Uniform steady state.}
To investigate the uniform equilibrium with an identically healthy population, we set $a = \abar$ producing the single ordinary differential equation\footnote{For $\gamma\le 1$, we set $\phisoc = 0$ before setting $a = \abar$ to avoid an undefined right-hand side of \eqref{eq:finalmodel}.},
\begin{equation} 
  \frac{\mathrm{d} a}{\mathrm{d} t} = 2 c \, (1-s) (\aopt-a).
  \label{unieq}
\end{equation}
The steady state (i.e., $\mathrm{d} a / \mathrm{d} t = 0$) is clearly $\displaystyle a=\aopt$. Linear stability analysis within this identical ornament manifold shows the fixed point $\displaystyle a=\aopt$ is stable for all $\gamma$, but numerical simulation suggests that the uniform fixed point is only stable for $\gamma \ge 2$. To resolve this apparent discrepancy, we investigate the uniform fixed point of \eqref{eq:finalmodel} in the continuum limit, and evaluate stability without restriction to the uniform manifold.  We are then able to find $\gamma$-dependence that agrees with simulations (details in supplementary information). 

%%%%%%%%%%%%%%%%%%%%%%%%%%%%%%%%%%%%%%%%%%%%%%
%                                                                                                                                             %
%                                                                TWO-MORPH FP                                                 %
%                                                                                                                                             %
%%%%%%%%%%%%%%%%%%%%%%%%%%%%%%%%%%%%%%%%%%%%%%

\paragraph{Two-morph steady state.}
To investigate the two-morph equilibrium, we assume all males have one of two ornament sizes $a_1$ and $a_2$. Taking $x$ to be the fraction of males with ornament size $a_1$, and $N \to \infty$, the dynamical system becomes
\begin{equation} 
  \begin{aligned}
  \frac{\mathrm{d} a_1}{\mathrm{d} t} &= \scriptstyle c \Big[ s \, \gamma \Big((1-x) |a_1 - a_2|\Big)^{\gamma-1} +  2 \, (1-s) (\aopt-a_1) \, \Big] \\
  \frac{\mathrm{d} a_2}{\mathrm{d} t} &= \scriptstyle c \Big[ s \, \gamma \Big(x |a_1 - a_2|\Big)^{\gamma-1} +  2 \, (1-s) (\aopt-a_2) \, \Big]~.
  \label{twoeq}
  \end{aligned} 
\end{equation}

\begin{figure}[htb!]
  \centering
    \includegraphics[width=\columnwidth]{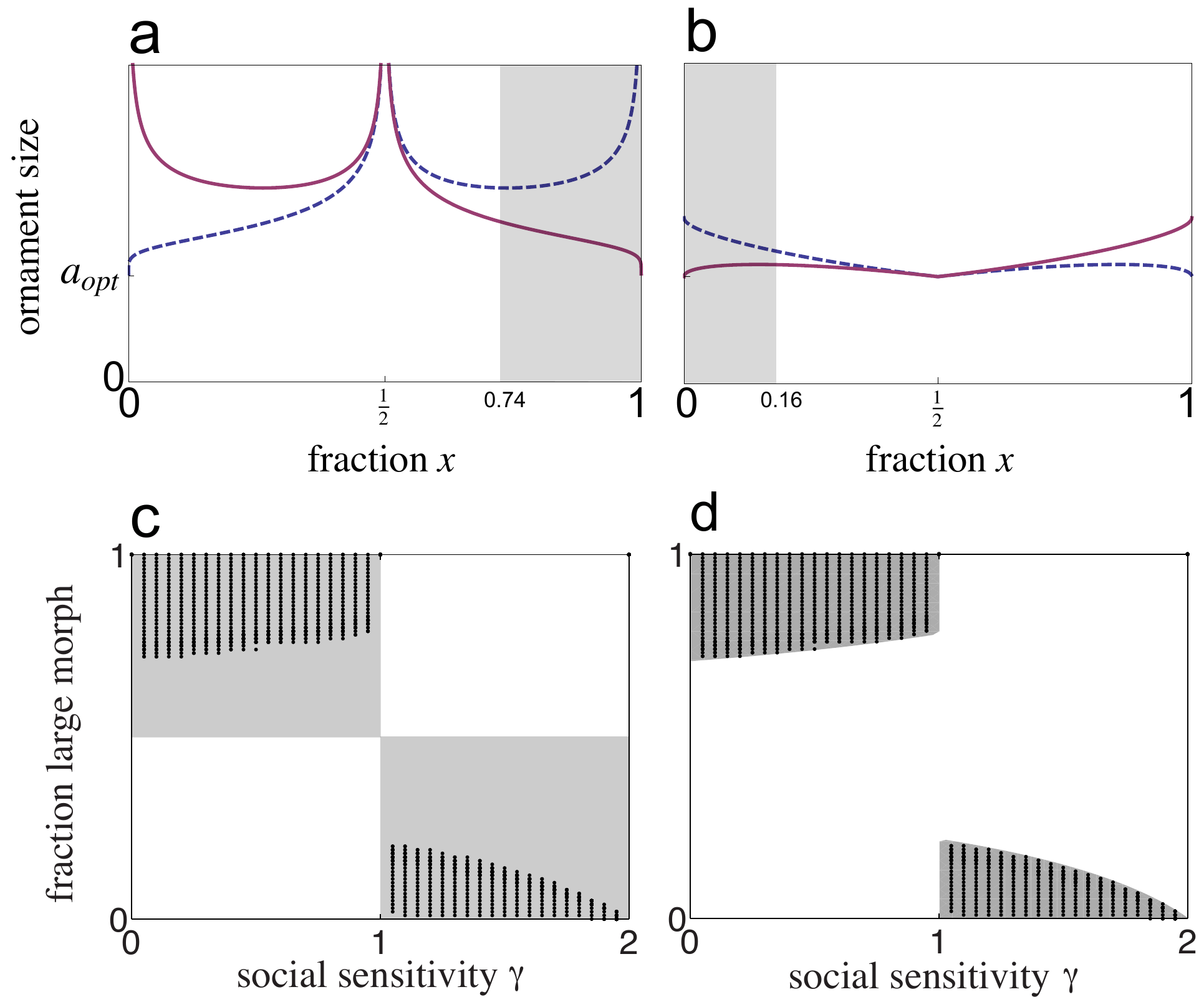}
  \caption{Stability regions for two-morph steady states ($N=100$, $s=1/2$). The ornament size for morph 1 is blue (dashed line), and the ornament size for morph 2 is maroon (solid line). The shaded regions are stable. \textbf{(a)} Two morph steady state for various morph fractionation $x$ and $\gamma = 0.5$  \textbf{(b)} Two morph steady state for various morph fractionation $x$ and $\gamma = 1.5$. \textbf{(c)} Analytical stability region (grey shading) for finite $N$ model within two-morph manifold with numerical stability region (dots) superimposed. \textbf{(d)} Analytical stability region (grey shading) from continuum model with numerical stability region (dots) superimposed.}
  \label{fig:twomorphsoln}
\end{figure}

There exists one two-morph steady state (i.e., solution to $\mathrm{d} a_1 / \mathrm{d} t = \mathrm{d} a_2 / \mathrm{d} t = 0$):
\begin{equation} 
  \begin{aligned} \scriptstyle
  a_1 =& \scriptstyle \, \aopt + \left( \frac{s \gamma}{2 (1-s)} \right)^{\frac{1}{2-\gamma}} \left( (1-x) \, \Big| \frac{(1-x)^\gamma x - x^\gamma +x^{1-\gamma}}{(1-x)x} \Big|^{\frac{1}{2-\gamma}} \right)^{\gamma-1} \\
  \scriptstyle a_2 =& \scriptstyle \, \aopt + \left( \frac{s \gamma}{2 (1-s)} \right)^{\frac{1}{2-\gamma}} \left( x \, \Big| \frac{(1-x)^\gamma x - x^\gamma +x^{1-\gamma}}{(1-x)x} \Big|^{\frac{1}{2-\gamma}} \right)^{\gamma-1}.
  \label{eq:twomorph}
  \end{aligned} 
\end{equation}
Figure~\ref{fig:twomorphsoln} (a),(b) shows how two-morph equilibria vary with the morph fractionation $x$. Within the shaded region, the fixed point is stable. To be clear, the model predicts that a bimodal population will emerge, with the fraction $x$ of the individuals within the population possessing ornaments of size $a_1$. We are \textit{not} claiming that a proportion $x$ of populations will evolve to ornament size $a_1$.

The eigenvalues for the linearized system constrained to this two-morph manifold are $\lambda_1 = -2(1-s)/s$ and $\lambda_2 = 2(\gamma-2)(1-s)/s$. Clearly, the two-morph equilibrium is stable (within the two-morph manifold) for $0 < \gamma < 2$ and unstable for $\gamma > 2$, when $\lambda_2>0$. Curiously, the stability of the two-morph equilibrium does not depend on $x$, the morph fractionation. This presents an apparent problem because numerical simulation suggests that only certain ranges of $x$ are stable: see figure~\ref{fig:twomorphsoln} (c). Similarly to the uniform fixed point analysis, we investigate the fixed points of the model in the continuum limit, and evaluate stability without restriction to any manifold.  We are then able to find $x$-dependence that agrees well with simulations: see figure~\ref{fig:twomorphsoln} (d) (details in supplementary information). 

%%%%%%%%% TWO MORPH EQUILIBRIA FIGURE %%%%%%%%%%%%
%%%%%%%%% TWO MORPH EQUILIBRIA FIGURE %%%%%%%%%%%%

%%%%%%%%%%%%%%%%%%%%%%%%%%%%%%%%%%%%%%%%%%%%%%%
%%                                                                                                                                             %
%%                                                                              DATA                                                      %
%%                                                                                                                                             %
%%%%%%%%%%%%%%%%%%%%%%%%%%%%%%%%%%%%%%%%%%%%%%%

\section{Model Validation\label{sec:data}}
We now revisit our simplifying assumption that all males are equally healthy. More realistically, we allow the intrinsic health $h$ to be taken from some distribution (perhaps set by genetic, developmental, or environmental factors). Suppose this distribution is such that the individual optimal ornament size $\aopt(h)$ is normally distributed. Then the stable two-morph steady state changes from a weighted sum of perfectly narrow Dirac delta functions to a distribution roughly resembling the sum of two Gaussians---usually a \textit{bimodal} distribution. Marginal histograms in figure~\ref{fig:fig1} (a),(b) show examples of steady states with varied intrinsic health. 

%As a reminder, our model is on a logarithmic scale, so we exponentiate all $a$ values to yield ornament sizes as measured in the real world. 

These examples resemble data from many species that grow ornaments. Figure~\ref{fig:fig1} (c),(d) show two examples of real-world ornament distributions that exhibit bimodality. Note that we do not expect the exact shape of the real-world distributions to match our simulations because the measured quantities will not necessarily be linear in cost. However, bimodality will be preserved regardless of the measured quantity. 

In a literature search \cite{Searcy:1990ta,Norris:1990uy,Bortolotti:2006co,BADYAEV:2000jy,Niecke:2003gi,Andersson:2002gn,Petrie:1994wp,Loyau:2005dc,MaysJr:2004vg,Pelabon:1998vo,West:2002cp,Moczek:2002vm,Barber:2001vm,Hyatt:1978tw,Tomkins:2005jf,Skarstein:1996uc}, we found a number of published data sets showing size distributions of suspected ornaments; 23 were of sufficient quality for testing agreement with this model. In 13 of those data sets we found some evidence for rejecting the hypothesis of unimodality: the data were more consistent with a mixture of two or more Gaussian distributions than with a single Gaussian. In seven data sets, we found stronger evidence: non-parametric tests rejected the hypothesis of unimodality. Note that other data sets were not inconsistent with bimodality, but small sample sizes often limited the power of statistical testing. See supplementary information for histograms and statistical tests of additional data sets.

%%%%%%%%%%%%%%%%%%%%%%%%%%%%%%%%%%%%%%%%%%%%%%
%                                                                                                                                             %
%                                                                 DISCUSSION                                                      %
%                                                                                                                                             %
%%%%%%%%%%%%%%%%%%%%%%%%%%%%%%%%%%%%%%%%%%%%%%

\section{Discussion \label{sec:discussion}}
\subsection{Implications for honest signaling}
Assuming this model adequately represents the handicap principle, we may ask if ornament size really does honestly advertise quality. In other words, if a female can choose among all the males, is she able to detect the healthiest (or weakest) males simply by looking at ornament size? Again taking the optimal ornament size $\aopt$ to be normally distributed, we examine the Kendall rank correlation between intrinsic health (as indicated by our proxy $\aopt$) and equilibrium ornament size. 

\begin{figure}[htb!]
  \begin{center}
    \includegraphics[width=\columnwidth]{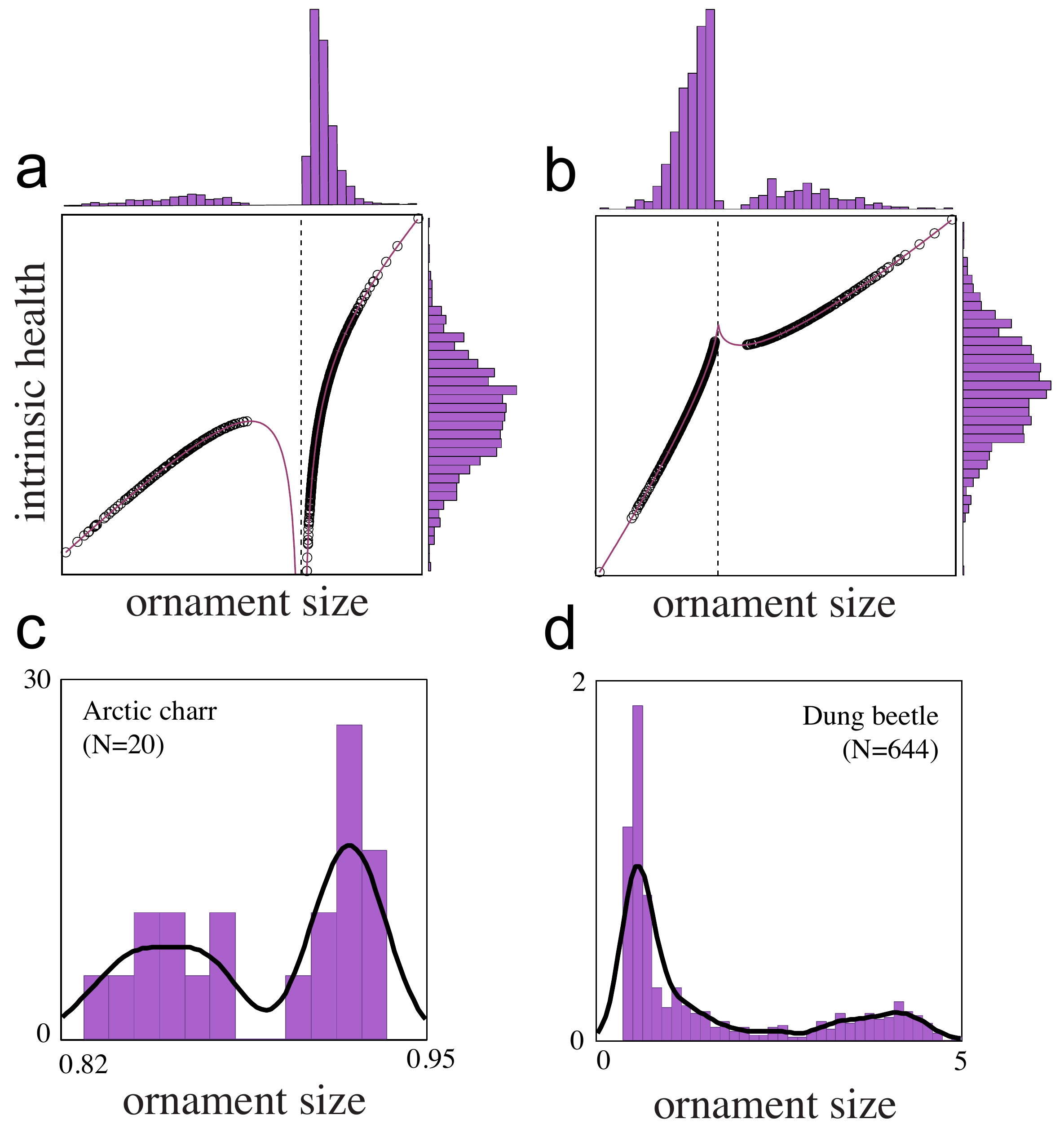}
  \end{center}
  \caption{Ornament size distributions in model and real-world data. Due to smaller sample sizes in real-world data, we superimpose a kernel density estimate (KDE) over the histograms as a visual aid (solid black line). \textbf{(a)} Simulation of model with $N=1000$ individuals, $\gamma=0.5$, $s=1/2$ (Kendall's rank correlation $\tau=0.9149$). \textbf{(b)}  Simulation of model with $N=1000$ individuals, $\gamma=1.5$, $s=1/2$ (Kendall's rank correlation $\tau=0.9998$).  In both (a) and (b), black dashed line ($a=\abar$) shows division between morphs, solid maroon curve shows analytical solution. Marginal histograms illustrate that normal distribution of $\aopt$ (proxy for intrinsic health) leads to bimodal distribution of $a$. \textbf{(c)}  Normalized histogram for arctic charr brightness \cite{Skarstein:1996uc} ($N$=20, KDE bandwidth=0.01).  \textbf{(d)} Normalized histogram for dung beetle horn length \cite{Moczek:2002vm} ($N$=644, KDE bandwidth=0.2).  }
  \label{fig:fig1}
\end{figure}

We find that the advertising is mostly honest, at least for large enough variance in health. Both observational and experimental work supports this finding \cite{Johnstone:1995wt}. Figure~\ref{fig:fig1} (a),(b) show examples of ornament size versus intrinsic health based on our model.

%%%%%%%%% DATA FIGURE %%%%%%%%%%%%
%%%%%%%%% DATA FIGURE %%%%%%%%%%%%

\subsection{Generality \label{sec:generality}}
It is natural to wonder about the generality of the results we have presented here. For a reasonable set of potential functions (described below), the only possible stable equilibria are multimodal distributions of ornament size. The following are the requirements for our reasonable potential functions:
\begin{enumerate}
\item Individual effects dominate potential for large ornament sizes. Specifically,
\[(1-s) \left| \frac{\partial}{\partial a} \phiind \right| > s \left| \frac{\partial}{\partial a} \phisoc \right| \text{ as } a\to\infty. \] 
This prevents ornament size from growing without bound, as can occur in model \eqref{eq:finalmodel} for $\gamma \ge 2$ (see figure~\ref{fig:model} (e)). 

\item Social effects dominate potential for at least some range of ornament sizes greater than the population mean. In other words, 
\[(1-s) \left| \frac{\partial}{\partial a} \phiind \right| < s \left| \frac{\partial}{\partial a} \phisoc \right|\] for at least some range of $a > \abar$. Failure to meet this criterion could be considered ``false" ornamentation, as in model \eqref{eq:finalmodel} for $\gamma= 1$. 

\end{enumerate} 
Assuming the potential functions are continuous\footnote{This is a stronger requirement than necessary. Actually, we only require that the two-sided limits exist everywhere.}, these criteria guarantee that two or more morphs will emerge (see supplementary information for details).

%One benefit of our modeling approach is that the evolutionarily optimal distribution we predict is independent of the particular mechanism(s) used to maintain phenotype diversity (which might include various combinations of genetic, epigenetic, environmental, or other cues - see supplementary information for more discussion of this point).  Of course, our model is not the only one that can show bimodality of traits. We have provided one simple mechanism for explaining such polymorphism, but other effects (e.g., intra-sexual selection, over-dominance, negative frequency-dependent selection) might also work in concert or be able to independently explain observed data. 

%%%%%%%%%%%%%%%%%%%%%%%%%%%%%%%%%%%%%%%%%%%%%%
%                                                                                                                                             %
%                                                                 CONCLUSION                                                    %
%                                                                                                                                             %
%%%%%%%%%%%%%%%%%%%%%%%%%%%%%%%%%%%%%%%%%%%%%%

\section{Conclusions \label{sec:conclude}}

The independent evolution of costly ornamentation across species has puzzled scientists for over a century. Several general evolutionary principles have been proposed to explain this phenomenon. Among the prominent hypotheses is the handicap principle, which posits that only the healthiest individuals can afford to grow and carry large ornaments, thereby serving as honest advertising to potential mates. We base a minimal model on this idea and find that, surprisingly, it predicts two-morph stratification of ornament size, which appears to be common in nature.

Importantly, the two morphs both have ornament sizes larger than the optimum for lone individuals. This means that the population survival potential, as indicated by the population average of individual potential $\displaystyle \overline{\varphi}^{(ind)}$, is reduced. Due to the presence of ornaments, we conclude that the evolutionary benefits of honest advertising must outweigh the net costs of ornamentation when the displays exist in nature.

%%%%%%%%%%%%%%%%%%%%%%%%%%%%%%%%%%%%%%%%%%%%%%
%
%          AUTHOR CONTRIBUTIONS
%
%%%%%%%%%%%%%%%%%%%%%%%%%%%%%%%%%%%%%%%%%%%%%%

\section{Author Contributions}
S.M.C.~and D.M.A.~developed and analyzed the model, S.M.C.~implemented the numerical simulations and created the database, R.I.B.~and S.M.C.~performed statistical tests on data. 
%All authors gave final approval for publication.

%%%%%%%%%%%%%%%%%%%%%%%%%%%%%%%%%%%%%%%%%%%%%%
%
%          DATA ACCESSIBILITY
%
%%%%%%%%%%%%%%%%%%%%%%%%%%%%%%%%%%%%%%%%%%%%%%

\section{Data Accessibility}
Data and code available from the Dryad Digital Repository \cite{dryadDataSoftware}: \url{http://dx.doi.org/10.5061/dryad.vb1pp}.

 %%%%%%%%%%%%%%%%%%%%%%%%%%%%%%%%%%%%%%%%%%%%%%
%
%          COMPETING INTERESTS
%
%%%%%%%%%%%%%%%%%%%%%%%%%%%%%%%%%%%%%%%%%%%%%%

\section{Competing Interests}
The authors have no competing interests.
 
%%%%%%%%%%%%%%%%%%%%%%%%%%%%%%%%%%%%%%%%%%%%%%
%
%          ACKNOWLEDGMENTS
%
%%%%%%%%%%%%%%%%%%%%%%%%%%%%%%%%%%%%%%%%%%%%%%

\section{Acknowledgments}
We thank Daniel Thomas, Trevor Price and Stephen Pruett-Jones for valuable discussions and Doug Emlen, Craig Packer, Markus Eichhorn, and Armin Moczek for sharing biological data. 

%and editors Innes Cuthill and Samuel Flaxman as well as anonymous referees for useful feedback.   

%%%%%%%%%%%%%%%%%%%%%%%%%%%%%%%%%%%%%%%%%%%%%%
%
%          FUNDING
%
%%%%%%%%%%%%%%%%%%%%%%%%%%%%%%%%%%%%%%%%%%%%%%

\section{Funding}
This research was supported in part by James S. McDonnell Foundation Grant No. 220020230 and National Science Foundation Graduate Research Fellowship No. DGE-1324585.

%%%%%%%%%%%%%%%%%%%%%%%%%%%%%%%%%%%%%%%%%%%%%%
%                                                                                                                                             
%               BIBLIOGRAPHY                                                         
%
%%%%%%%%%%%%%%%%%%%%%%%%%%%%%%%%%%%%%%%%%%%%%%

\bibliographystyle{ieeetr}
\bibliography{ornamentLibraryV11}

\clearpage
\newpage



\section*{Supplementary material}

# Supplemental: Handicap principle implies emergence of dimorphic ornaments

Sara M. Clifton,[1, *] Rosemary I. Braun,[2, 3] and Daniel M. Abrams[1, 3, 4]

[1]*Department of Engineering Sciences and Applied Mathematics,*
*Northwestern University, Evanston, Illinois 60208, USA*
[2]*Division of Biostatistics, Northwestern University, Chicago, Illinois 60611, USA*
[3]*Northwestern Institute for Complex Systems, Northwestern University, Evanston, Illinois 60208, USA*
[4]*Department of Physics and Astronomy, Northwestern University, Evanston, Illinois 60208, USA*

## I. THE CONNECTION BETWEEN POTENTIAL AND FITNESS

Many evolutionary dynamics problems begin with the replicator equation [1], which in the continuum limit is as follows:

$$\frac{\partial p}{\partial t} = p(a, t) \left[ f(a, p) - \bar{f}(p) \right],\tag{S1}$$

where $p$ is the probability distribution of a continuous phenotype $a$ at time $t$, $f$ is the fitness of a phenotype (say, ornament size or brightness) given a population state, and $\bar{f} = \int_{-\infty}^{\infty} f(a, p) p(a, t)\, \mathrm{d}a$ is the average population fitness [2].

Given that probability must be conserved, the distribution of phenotypes must also follow the continuity equation

$$\frac{\partial p}{\partial t} = -\frac{\partial}{\partial a} \left( p \frac{\mathrm{d}a}{\mathrm{d}t} \right).\tag{S2}$$

This formulation differs from the replicator equation (S1) in that it requires specification of the phenotype flux $\mathrm{d}a/\mathrm{d}t$ rather than fitness $f$. Our approach treats this flux as derivable from some potential function, which we refer to as $\varphi$, the net "reproductive potential" (see equation (3)).

Intuitively, the relationship between our phenotype flux $\mathrm{d}a/\mathrm{d}t$ and the more commonly used replicator equation approach (the upward distribution flux) can be seen in figure S1. These reflect interchangeable ways of viewing the evolutionary process of optimizing the probability distribution $p(a, t)$.

We can express the relationship between the two approaches mathematically simply by equating the right-hand-sides of equations (S1) and (S2), yielding

$$f - \bar{f} = -\frac{1}{p} \frac{\partial}{\partial a} \left( p \frac{\mathrm{d}a}{\mathrm{d}t} \right) = -c \left( \frac{1}{p} \frac{\partial p}{\partial a} \frac{\partial \varphi}{\partial a} + \frac{\partial^2 \varphi}{\partial a^2} \right)\tag{S3}$$

where the last equality makes use of equation (4). Integrating equation (S3) once with respect to $a$ and using equation (4) yields an integro-differential equation for $\varphi$ in terms of $f$:

$$\frac{\partial \varphi}{\partial a} = -\frac{1}{cp} \int_{-\infty}^{a} p \left( f - \bar{f} \right) \mathrm{d}a,\tag{S4}$$

assuming $p\, \mathrm{d}a/\mathrm{d}t \to 0$ as $a \to -\infty$.

## II. FIXED POINTS FOR GENERAL CLASS OF POTENTIAL FUNCTIONS WITH MINIMUM OF 2 MORPHS

In our analysis of (5), we claim that "reasonable" potential functions lead to stratification from a nearly uniform population into multiple distinct morphs. Here we examine in more detail what we mean by "reasonable". Again we consider a potential function

$$\varphi = s\, \varphi^{(\mathrm{soc})} + (1 - s)\, \varphi^{(\mathrm{ind})}, \quad s \in [0, 1]$$

---

* E-mail me at: sclifton@u.northwestern.edu

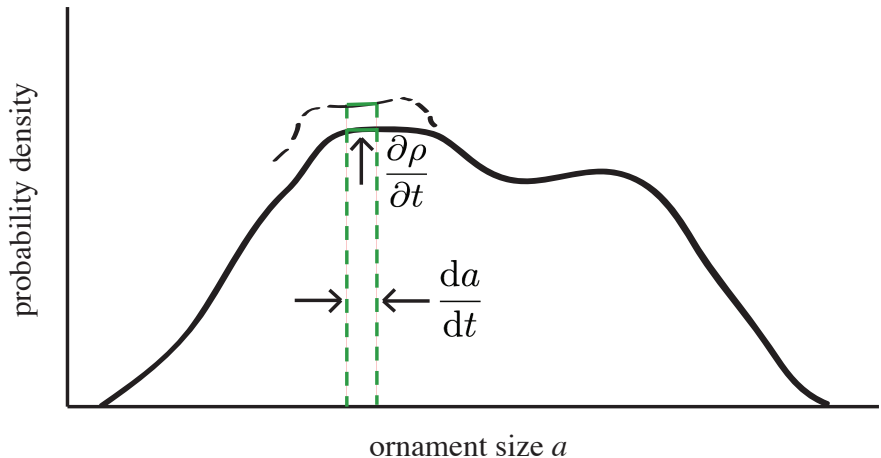

FIG. S1. (Color online) Consider an infinitesimal sliver (dashed green) of the probability density function at a particular time (solid black). After an infinitesimal time increment, the probability density function changes a small amount (dashed black). Because probability is conserved, the flux $da/dt$ of population ornament sizes into (or out of) the sliver increases (or decreases) the height of the probability density function.

where $\varphi^{(\mathrm{soc})}$ is a continuous and differentiable increasing function of ornament size, and $\varphi^{(\mathrm{ind})}$ is a continuous, singly-peaked function of ornament size. Assuming that the dynamics are such that ornaments grow on an evolutionary time scale at a rate proportional to marginal potential gain,

$$\frac{\mathrm{d}a}{\mathrm{d}t} \propto \frac{\partial}{\partial a}\varphi,$$

and we have $\dfrac{\mathrm{d}a}{\mathrm{d}t} = 0$ only for $a \geq a_{\mathrm{opt}}$.

We further assume that the following two criteria are satisfied:

1. Individual effects dominate reproductive potential for large ornament sizes. Specifically,

$$(1-s)\left|\frac{\partial}{\partial a}\varphi^{(\mathrm{ind})}\right| > s\left|\frac{\partial}{\partial a}\varphi^{(\mathrm{soc})}\right| \text{ as } a \to \infty. \tag{S5}$$

   This prevents ornament size from growing without bound, as can occur in equation (5) for $\gamma \geq 2$.

2. Social effects dominate reproductive potential for at least some range of ornament sizes greater than the population mean. In other words,

$$(1-s)\left|\frac{\partial}{\partial a}\varphi^{(\mathrm{ind})}\right| < s\left|\frac{\partial}{\partial a}\varphi^{(\mathrm{soc})}\right| \tag{S6}$$

   for at least some range of $a > \bar{a}$. Failure to meet this criterion could be considered "false" ornamentation, as can occur in equation (5) for $\gamma = 1$.

Assuming that the two-sided limits exist everywhere for both potential functions (a less strict requirement than continuity), these criteria guarantee that two or more morphs will emerge. See figure S2 (a)-(c) for graphical proof.

## III. FIXED POINTS FOR GENERAL CLASS OF POTENTIAL FUNCTIONS WITH MAXIMUM OF 2 MORPHS

In our fixed points analysis of equation (5), we only consider uniform and two-morph steady states. We now show that these are the only types of fixed points for a wider class of potential functions, including our potential function (3). Consider a more general total potential function

$$\varphi = s\,\varphi^{(\mathrm{soc})} + (1-s)\,\varphi^{(\mathrm{ind})}, \quad s \in [0,1] \tag{S7}$$

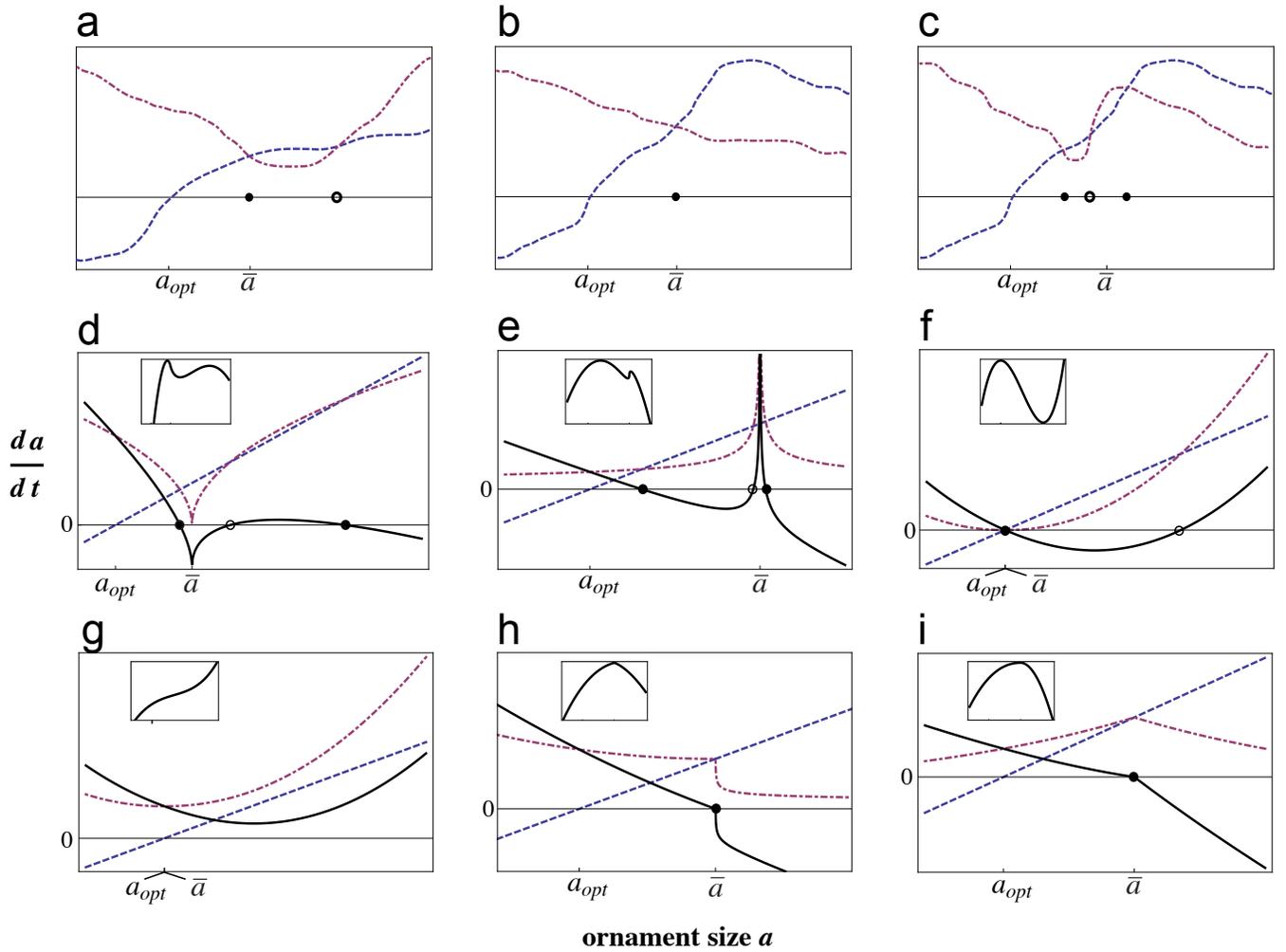

FIG. S2. (Color online) Sketched examples of derivatives of negated individual potential (dashed blue) and social potential (dot dashed maroon) for a single male in a population near equilibrium. The derivative of total potential (solid black) is proportional to $\mathrm{d}a/\mathrm{d}t$, so intersections of individual and social potentials are the fixed points. Stable fixed points are marked with a filled black dot, and unstable fixed points are marked with an unfilled black dot. The total potential is inset. **(a)** An example of potential functions that satisfy restriction (S6), but not restriction (S5). In this case, both a stable uniform state and unbounded growth are possible. **(b)** An example of potential functions that satisfy restriction (S5), but not restriction (S6). In this case, the population will evolve to a uniform state. **(c)** An example of potential functions satisfying both restrictions (S5) and (S6). These conditions guarantee that the population will evolve into at least two morphs. **(d)-(i)** With restrictions (S8), the only possible stable steady states (filled black dots) are one- or two-morphs. Note that the system may or may not have an unstable node (unfilled black dots), or it may have no fixed points.

where $\varphi^{(\mathrm{soc})}$ is a continuous and differentiable increasing function of ornament size, and $\varphi^{(\mathrm{ind})}$ is a continuous, singly-peaked function of ornament size. Similar to our previous general class of potential functions,

$$\frac{\mathrm{d}a}{\mathrm{d}t} \propto \frac{\partial}{\partial a}\varphi,$$

we conclude that $\dfrac{\mathrm{d}a}{\mathrm{d}t} = 0$ only for $a \geq a_{\mathrm{opt}}$. This implies that equilibrium ornament sizes (if an equilibrium exists) will all be at least as large as the optimal. Because this is a first order ordinary differential equation model, we also know that oscillations are not possible.

We further assume that

$$\frac{\partial^3}{\partial a^3} \varphi^{(\mathrm{ind})} \equiv 0$$

$$\frac{\partial^3}{\partial a^3} \varphi^{(\mathrm{soc})} > 0 \text{ or } \frac{\partial^3}{\partial a^3} \varphi^{(\mathrm{soc})} < 0, \quad a \neq \bar{a}. \tag{S8}$$

In other words, individual potential is quadratic, and the derivative of social potential is either concave up or concave down, except possibly at the mean. With these additional restrictions on the potential function, only uniform and two-morph stable fixed points are possible. See figure S2 (d)-(i) for graphical proof. Our model (5) satisfies all restrictions, so we conclude that our exploration of the one- and two-morph steady states was a thorough investigation of all possible fixed points.

## IV. CONTINUUM LIMIT

In the main manuscript, we derived the phenotype dynamics (5) of a system of $N$ population representatives. The fixed points of this system are a discrete set of ornament sizes. Now we take $N \to \infty$, which turns the $N$ ordinary differential equations into a partial integro-differential equation for a continuous distribution of ornament sizes $p(a, t)$. The equation we derive is the replicator function for continuous phenotypes [2].

We use conservation of probability to find the governing equation for the probability density function $p(a, t)$. The probability of a male having an ornament size in $(a, a + \mathrm{d}a)$ for small $\mathrm{d}a$ is approximately $p(a, t)\,\mathrm{d}a$. Given our assumption that individuals are neither created nor destroyed in $(a, a + \mathrm{d}a)$, we have

$$\frac{\partial p}{\partial t}\,\mathrm{d}a = p\left.\frac{\mathrm{d}a}{\mathrm{d}t}\right|_a - p\left.\frac{\mathrm{d}a}{\mathrm{d}t}\right|_{a+\mathrm{d}a}.$$

In other words, the change in individuals in the sliver $(a, a + \mathrm{d}a)$ is equal to the number that enter the sliver minus the number that leave. In the limit $\mathrm{d}a \to 0$, we get the continuity equation

$$\frac{\partial p}{\partial t} = -\frac{\partial}{\partial a}\left(p\,\frac{\mathrm{d}a}{\mathrm{d}t}\right). \tag{S9}$$

The dynamics of $a$ follow (5) in the limit $N \to \infty$

$$\frac{\mathrm{d}a}{\mathrm{d}t} = c\left[s\,\gamma|a - \bar{a}|^{\gamma-1} + 2(1-s)\,(a_{\mathrm{opt}} - a)\right], \tag{S10}$$

where the mean ornament size is

$$\bar{a} = \int_{-\infty}^{\infty} a(t)\,p(a, t)\,\mathrm{d}a.$$

We substitute (S10) into (S2) to get a partial integro-differential equation for the probability density function $p(a, t)$ for ornament size

$$\frac{\partial p}{\partial t} = -c\frac{\partial}{\partial a}\left(p\left[s\,\gamma|a - \bar{a}|^{\gamma-1} + 2(1-s)\,(a_{\mathrm{opt}} - a)\right]\right). \tag{S11}$$

### A. Continuum limit uniform steady state

Now that we have established the continuum limit of the discrete model, we wish to investigate the fixed points we found previously. Within this continuum framework, the uniform fixed point $a = a_{\mathrm{opt}}$ is the delta distribution

$$p(a, t) = \delta(a - a_{\mathrm{opt}}). \tag{S12}$$

Previously, we investigated the stability of the uniform steady state by perturbing every member of the population by the same arbitrary, small amount. If we wished to repeat this investigation for the continuum model, we would shift the peak of the delta function by an arbitrary small amount from $a_{\mathrm{opt}}$ to some $a_0$. To make stability analysis

more general, we also consider widening the delta function into a narrow Gaussian with an arbitrary small standard deviation $\sigma$. Figure S3 (a),(b) illustrate this idea.

We now wish to confirm that this continuum representation (S10) of the model is consistent with our discrete model (5), at least near the simplest fixed point (the uniform state). Based on our previous stability analysis, we expect that $a_0$ will shift back to $a_{\text{opt}}$ and the width of the peak will shrink to 0 for $\gamma \geq 2$. However, we do not know how quickly these shifts occur relative to each other.

We will first investigate the dynamics of $a_0$ (i.e. $\sigma$ is effectively constant on the time scale of interest). Then the "perturbed" distribution is the narrow Gaussian

$$p(a,t) = \frac{1}{\sigma\sqrt{2\pi}} e^{-(a-a_0(t))^2/2\sigma^2} \tag{S13}$$

with constant $\sigma \ll 1$ and $a_0(t)$ near the fixed point $a_{\text{opt}}$.

Plugging (S13) into the continuity equation (S2), and solving for the highest order (fastest) dynamics of $a_0$, we see

$$\frac{\mathrm{d}a_0}{\mathrm{d}t} = s\gamma|a-a_0|^{\gamma-1} + 2(1-s)(a_{\text{opt}} - a). \tag{S14}$$

Note that (S14) is only true if $\sigma \to 0^+$ faster than $a \to a_0$. If we instead assume $\sigma \to 0^+$ slower than $a \to a_0$, the right-hand side of (S14) is unbounded, and therefore inconsistent with the discrete model. Taking $a \to a_0$ in (S14), we see as expected

$$\frac{\mathrm{d}a_0}{\mathrm{d}t} = 2(1-s)(a_{\text{opt}} - a_0).$$

As we see that $\sigma$ shrinks to 0 faster than $a \to a_0$, we investigate the dynamics of $\sigma(t) \ll 1$ for $a_0 = a_{\text{opt}}$. Again, we take $p(a,t)$ to be a narrow Gaussian distribution

$$p(a,t) = \frac{1}{\sigma(t)\sqrt{2\pi}} e^{-(a-a_0)^2/2\sigma(t)^2}. \tag{S15}$$

Substituting (S15) into (S2) and Taylor expanding about $\sigma = 0$ gives

$$\frac{\mathrm{d}\sigma}{\mathrm{d}t} = \left[ \frac{\gamma|a-a_0|^{\gamma-1}}{a-a_0} + 2\frac{1-s}{s}\frac{a_{\text{opt}}-a}{a-a_0} \right]\sigma + \mathcal{O}(\sigma^3).$$

We see that as $a \to a_0 = a_{\text{opt}}$ for $\gamma < 2$, the uniform fixed point is unstable (coefficient of $\sigma$ is $\infty$). For $\gamma > 2$, the fixed point is stable (coefficient of $\sigma$ is $-2\dfrac{1-s}{s}$). The fixed point for $\gamma = 2$ is conditionally stable (coefficient of $\sigma$ is $\pm 2 - 2\dfrac{1-s}{s}$). These results agree with the finite $N$ model.

### B. Continuum limit two-morph steady state

Next, we investigate the stability of the two-morph steady state. Similar to our investigation of the uniform steady state, we "perturb" the two-morph steady state to the weighted sum of two narrow Gaussian distributions

$$p(a,t) = \frac{x}{\sigma_1(t)\sqrt{2\pi}} e^{-(a-a_1)^2/2\sigma_1(t)^2} + \frac{1-x}{\sigma_2(t)\sqrt{2\pi}} e^{-(a-a_2)^2/2\sigma_2(t)^2}, \tag{S16}$$

where $a_1$ and $a_2$ are given by the two-morph fixed point (8). Figure S3 (c),(d) illustrate this idea.

Plugging (S16) into the continuity equation (S2) and using $\bar{a} = xa_1 + (1-x)a_2$, we get a system of two ordinary differential equations for the evolution of $\sigma_1$ and $\sigma_2$:

$$\begin{aligned} \frac{\mathrm{d}\sigma_1}{\mathrm{d}t} &= \lambda_1\sigma_1 + \mathcal{O}(\sigma_1^3) \\ \frac{\mathrm{d}\sigma_2}{\mathrm{d}t} &= \lambda_2\sigma_2 + \mathcal{O}(\sigma_2^3), \end{aligned} \tag{S17}$$

where $\lambda_1$ and $\lambda_2$ depend on $a_{\text{opt}}, s, x,$ and $\gamma$ (expressions omitted due to length). Setting $a_{\text{opt}} = 1$ and $s = 1/2$ for instance, we plot the stability region (i.e., where $\lambda_1, \lambda_2 < 0$) for the two-morph steady state in terms of social sensitivity $\gamma$ and the proportion of males in the large-ornamented group. See figure 3 (d). This is the same stability region we found numerically, which resolves the apparent discrepancy we saw when perturbing the locations of the peaks, but not the widths of the peaks of the two-morph steady state distribution. We have confirmed numerically that convergence to the two-morph fixed points is approximately exponential.

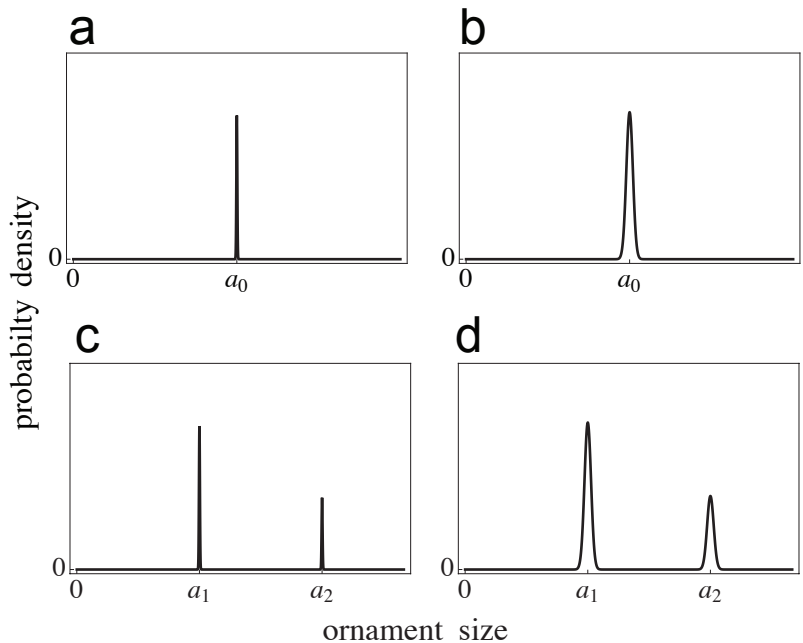

FIG. S3. We consider perturbations to the uniform fixed point $a = a_{\mathrm{opt}}$ and the two-morph fixed point in equation (8) such that the peaks of the distribution are centered at the fixed point solution, and the widths of the peaks are nearly 0. **(a)** Shift peak of the delta uniform solution to $a_0$. **(b)** Perturb peak width of the delta uniform solution. **(c)** Two-morph steady state. **(d)** Perturb peak widths of the delta two-morph solution.

## V. EIGENVALUES OF SYSTEM AS $N \to \infty$

When investigating the stability of the two morph steady state, we chose to take the continuum limit of the model and then investigate the dynamics of the standard deviation of a Gaussian perturbation to the two morph equilibrium. Now we look at the eigenvalues of the finite $N$ system in the limit $N \to \infty$.

Scaling time such that $c = 1$, the Jacobian for the system (5) has diagonal elements

$$J_{ii} = s\gamma(\gamma - 1)\left(1 - \frac{1}{N}\right)^2 \mathrm{sgn}\,(a_i - \bar{a})|a_i - \bar{a}|^{\gamma - 2} - 2(1 - s),$$

and off-diagonal elements

$$J_{ij} = s\gamma(\gamma - 1)\left(-\frac{1}{N}\right)\left(1 - \frac{1}{N}\right)\mathrm{sgn}\,(a_i - \bar{a})|a_i - \bar{a}|^{\gamma - 2}.$$

As $N \to \infty$,

$$J_{ii} \to s\gamma(\gamma - 1)\mathrm{sgn}\,(a_i - \bar{a})|a_i - \bar{a}|^{\gamma - 2} - 2(1 - s)$$
$$J_{ij} \to 0,$$

indicating that for large $N$, the Jacobian matrix is approximately diagonal. Therefore, the diagonal elements are approximately the eigenvalues. Plugging in the two morph fixed point (8), we get two eigenvalues $\lambda_1$ and $\lambda_2$ with multiplicity $xN$ and $(1 - x)N$ respectively. If we plot the stability region (i.e. where $\lambda_1, \lambda_2 < 0$), we see that it's the same as that of the continuum model seen in figure 3 (d).

## VI. STATISTICAL ANALYSIS OF ORNAMENTATION DATA

Our model for the evolution of costly mating displays predicts the emergence of two distinct morphs of ornament sizes. We tested whether the two-morph state was detectable in a variety of ornament datasets (figures S5,S6). Three

approaches were used: a parametric mixture–model fit; the nonparametric but highly conservative Hartigans' Dip Test for bimodality [5]; and a simulation–based nonparametric test which improves upon the Hartigan test sensitivity.

We present test results for Hartigans' Dip Test and the simulation–based nonparametric test, called the LUU (Least Unimodal Unimodal) test for reasons that will be clear, in table I. Test results for the parametric–model fit are in table II.

### A. Parametric two-morph test

All count and size measurement data were log-transformed prior to analysis (as is typical for physical measurements) to account for the bounded support of size distributions. Here, we make the assumption that ornament sizes within a morph will be log-normally distributed, and that a multi-morph state will exhibit a mixture of distributions. We thus fit Gaussian mixture models with 1–5 components of unequal variance to the log-transformed data and find the number of components that yields the best BIC [6]. In the absence of a social fitness pressure, we expect the best fit to be a single Gaussian (corresponding to the one morph state), while the two–morph state predicted from our model will have the best fit with $\geq 2$ components.

### B. Hartigans' dip test

An essential drawback of using the above mixture model fit to assess the number of ornament–size morphs in the data is that it is extremely sensitive to deviations from the parametric assumption that a one–morph state will be well–described by a single Gaussian. False positives are likely when those assumptions are violated; if a single–morph state has a skewed (or otherwise non-normal) distribution, a mixture of $\geq 2$ Gaussians will generally give a higher BIC than a single–component distribution.

A more conservative approach is to look for evidence of strict multi-modality (with dips in the distribution), rather than a mixture (which may not exhibit a "dip"). Hartigan and Hartigan define the dip statistic $D$ as the maximum difference between the empirical cumulative distribution function and the CDF of the unimodal distribution that minimizes that maximum difference. The reference distribution is customarily taken to be the uniform distribution, the least singly–peaked of all unimodal distributions. The $p$-value for the dip is calculated by comparing $D$ to those obtained from repeated samples of the same size drawn from a uniform distribution. The dip test thus measures whether the empirical distribution of the data exhibits greater departure from unimodality than would be expected from a sample of the same size if the underlying distribution were uniform.

### C. Bootstrap dip test

While the mixture test may be overly sensitive in detecting deviations from a single morph, Hartigans' dip test is likely to be excessively conservative and insensitive at small sample sizes. A finite sample drawn from a uniform distribution will, with high probability, have a larger dip by chance than a finite sample drawn from a two–morph distribution such as those shown in figure 4 (a),(b).

To address this problem, we propose a bootstrap dip test which takes as its reference distribution the "least unimodal" unimodal density estimate of the sample. Given a finite sample, we construct a kernel density estimate (KDE) using a Gaussian kernel at various bandwidths. At very large bandwidths, the KDE will be unimodal; as the bandwidth is reduced, the KDE will approach a multimodal distribution with as many modes as there are unique values in the dataset. We define the least–unimodal unimodal (LUU) distribution to be that obtained from the smallest bandwidth for which the KDE is still strictly unimodal.

From this LUU density estimate, we generate random samples of the same size as the original data, and compute their dip statistics. These bootstrapped samples serve as the reference distribution against which the dip statistic of the data is compared. This test thus measures whether the empirical distribution of the data exhibits greater departure from unimodality than would be expected from a sample of the same size if the underlying distribution were *the unimodal distribution best fit to the sample.* Figure S4 illustrates that this bootstrap dip test is more sensitive to bimodality than Hartigans' Dip Test.

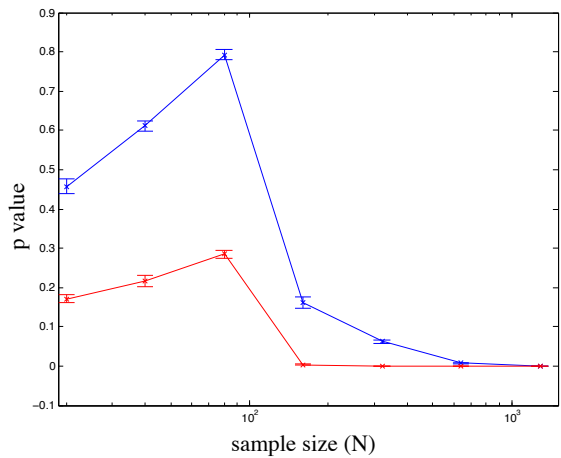

FIG. S4. For small sample sizes of bimodal data, like we have for most of our animal data sets, the p-values for bimodality using Hartigans' Dip Test (blue) are larger than our bootstrap dip test (red). As the sample size increases, we gain significance using our test first and Hartigans' Dip Test eventually, showing our test is less conservative. The data used here are equilibrium states of our model (5) for $\gamma = 1.5$, $s = 0.5$, and $a_{\text{opt}}$ drawn from a normal distribution with mean 1 and standard deviation 0.25. We know these samples are bimodal. Error bars are standard deviations from 10 trials.

## VII. ADDITIONAL DATA AND ANALYSIS

We have additional data sets of ornament distribution from various species in figures S5 and S6. The kernel density curves are superimposed for reference. If body size is a form of advertising, then we may also use data of salmon [25], trout [26], wolf spiders [27], and other bimodally distributed species. See figure S7.

While this work is based on mating displays in the animal kingdom, we hypothesize that similar forces operate on plants that compete within their own species for resources. For instance, a tree's height could be analogous to ornament size in our model, in that growing taller incurs costs to the individual, but being relatively taller in a forest has competitive benefits. In fact, certain tree species exhibit bimodal height distributions [28, 29]. See figure S8.

## VIII. CONNECTION TO SPECIATION MODELS

We speculate that the mechanism we describe here may also have implications for speciation. Models of speciation presented in Lande [4] and Stewart [31] are similar to our ornamentation model in both form and outcome. Stewart claims that for an all-to-all system of behaviorally identical individuals (like ours), the population will split into two species for most environmental conditions. Like our social sensitivity $\gamma$, Stewart's environmental factor $\lambda$ varies on a slow time scale relative to the dynamical system. Also like our model, Stewart's model exhibits similar fractionation (simulating 100 individuals, the population splits into "clumps" of 84 and 16).

Lande uses quantitative genetics techniques to show that sexual selection may lead to speciation. Our model is quite similar to Lande's model interpreted on a logarithmic scale. Like our model, Lande's sexual selection alone would lead to runaway ornament sizes, but natural selection stabilizes growth. Unlike our model, Lande states that "natural selection on mating preferences also creates the possibility of evolutionary oscillations." Because we ignore the long time scale effects of female choice, our model precludes the possibility of oscillations.

| Data set | N | p-value (Dip test) | p-value (LUU test) | p-value (Dip test - log data) | p-value (LUU test - log data) | Tests reject unimodality? |
|---|---|---|---|---|---|---|
| Dung beetle horn length (Emlen [7]) | 223 | 0.0011** | 0.0001*** | 0.0035** | 0.0000*** | yes |
| Yellow-breasted chat plumage coloration (Mays [8]) | 62 | 0.1932 | 0.0530 | 0.5479 | 0.2652 | no |
| Peacock eye spots (Loyau [9]) | 24 | 0.6390 | 0.3793 | 0.5965 | 0.3187 | no |
| Peacock eye spots (Petrie [10]) | 24 | 0.9183 | 0.7682 | 0.8809 | 0.6963 | no |
| Peacock eye spots (Loyau/Petrie merged) | 48 | 0.9016 | 0.6699 | 0.9006 | 0.6587 | no |
| Arctic charr skin brightness (Skarstein [11]) | 20 | 0.2633 | 0.1558 | 0.2802 | 0.1658 | no |
| Salmon body size (Glover [12]) | 72 | 0.6206 | 0.1467 | 0.7432 | 0.2497 | no |
| Widowbird tail length (Anderson [13]) | 107 | 0.9992 | 0.9700 | 0.9972 | 0.9594 | no |
| Widowbird red collar patch size (Anderson [13]) | 107 | 0.0046** | 0.0002*** | 0.0317* | 0.0030** | yes |
| Barn owl spottiness (Nieche [14]) | 20 | 0.6476 | 0.3858 | 0.7196 | 0.5157 | no |
| Finch carotenoid coloration (Badyaev [15]) | 68 | 0.5295 | 0.1927 | NA | NA | no |
| Stickleback nest compactness (Barber [16]) | 38 | 0.6085 | 0.2221 | NA | NA | no |
| Partridge black ventral area (Bortolotti [17]) | 29 | 0.9032 | 0.6652 | 0.8704 | 0.5812 | no |
| Roe deer antler length (Pelabon [18]) | 242 | 0.0341* | 0.0012** | 0.0232* | 0.0001*** | yes |
| Lion >2.2 yrs mane length (West [19]) | 441 | 0.8687 | 0.4134 | 0.9873 | 0.9521 | no |
| Lion >2.2 yrs mane darkness (West [19]) | 442 | 0.9078 | 0.6698 | 0.9602 | 0.9033 | no |
| Lion >5 yrs mane length (West [19]) | 257 | 0.8085 | 0.4779 | 0.8557 | 0.5356 | no |
| Lion >5 yrs mane darkness (West [19]) | 257 | 0.8285 | 0.4129 | 0.8567 | 0.5173 | no |
| Dung beetle horn length - WA (Moczek [20]) | 644 | 0.0000*** | 0.0000*** | 0.0000*** | 0.0000*** | yes |
| Dung beetle horn length - NC (Moczek [20]) | 1016 | 0.0000*** | 0.0000*** | 0.0000*** | 0.0000*** | yes |
| Earwig forceps length (Tomkins [21]) | 134 | 0.0000*** | 0.0000*** | 0.0000*** | 0.0000*** | yes |
| Great tit stripe length (Norris [22]) | 63 | 0.2034 | 0.0781 | NA | NA | no |
| Fiddler crab fight duration (Hyatt [23]) | 80 | 0.7059 | 0.2601 | 0.6362 | 0.3312 | no |
| Fiddler crab fight acts (Hyatt [23]) | 80 | 0.8966 | 0.5273 | 0.9006 | 0.5714 | no |

TABLE I. Unimodality test results for animal ornamentation data sets. Hartigans' Dip Test (Dip test) is more conservative than our bootstrap dip test (LUU test); therefore our LUU test is more likely to reject unimodality. We performed both tests on log-transformed data because tissue measurements are often log-normally distributed [24]. We note in the rightmost column if the unimodality tests reject the null hypothesis that the distributions of ornament size are unimodal. Note that we exclude p-values for log-transformed data (NA) if the original data is not a straight-forward measurement of tissue investment.

| Data set | N | fractionation | morph means | morph variances | fractionation (log data) | morph means (log data) | morph variances (log data) |
|---|---|---|---|---|---|---|---|
| Dung beetle horn length (Emlen [7]) | 223 | 0.2372 0.2677 0.2414 0.2156 0.0380 | 0.2631 1.0576 0.7280 0.1204 0.5126 | 0.0055 0.0112 0.0142 0.0018 0.0000 | 0.0448 0.2103 0.3299 0.1950 0.2200 | -2.8934 -1.3094 -0.3286 -2.0101 0.0629 | 0.0005 0.0509 0.0553 0.0412 0.0082 |
| Yellow-breasted chat plumage coloration (Mays [8]) | 62 | 0.7247 0.2753 | 40.2987 23.7743 | 58.3743 4.9154 | 0.2924 0.7076 | 3.1794 3.6963 | 0.0084 0.0302 |
| Peacock eye spots (Loyau [9]) | 24 | 1.0000 | 152.0645 | 46.7236 | 1.0000 | 5.0233 | 0.0021 |
| Peacock eye spots (Petrie [10]) | 24 | 1.0000 | 145.9515 | 95.9004 | 1.0000 | 4.981 | 0.0046 |
| Peacock eye spots (Loyau/Petrie merged) | 48 | 1.0000 | 149.0080 | 80.6543 | 1.0000 | 5.0021 | 0.0038 |
| Arctic charr skin brightness (Skarstein [11]) | 20 | 0.4505 0.5495 | 2.3538 2.5160 | 0.0015 0.0004 | 0.4507 0.5493 | 0.8559 0.9226 | 0.0003 0.0001 |
| Salmon body size (Glover [12]) | 72 | 0.1383 0.8617 | 9.3169 14.6375 | 0.5107 1.0055 | 0.1388 0.8612 | 2.2296 2.6814 | 0.0056 0.0046 |
| Widowbird tail length (Anderson [13]) | 107 | 1.0000 | 221.5356 | 796.5005 | 1.0000 | 5.3920 | 0.0179 |
| Widowbird red collar patch size (Anderson [13]) | 107 | 1.0000 | 222.1704 | 2419.6 | 1.0000 | 5.3779 | 0.0526 |
| Barn owl spottiness (Nieche [14]) | 20 | 1.0000 | 1.2436 | 0.4555 | 1.0000 | 0.0695 | 0.3068 |
| Finch carotenoid coloration (Badyaev [15]) | 68 | 1.0000 | 1.7732 | 3.1678 | NA | NA | NA |
| Stickleback nest compactness (Barber [16]) | 38 | 0.8947 0.1053 | 37.7314 90.2335 | 99.9319 0.0131 | NA | NA | NA |
| Partridge black ventral area (Bortolotti [17]) | 29 | 1.0000 | 21.1812 | 56.7020 | 1.0000 | 2.9779 | 0.1728 |
| Roe deer antler length (Pelabon [18]) | 242 | 0.0903 0.9097 | 12.1801 18.1135 | 7.7178 5.1123 | 0.1235 0.8765 | 2.5521 2.8933 | 0.0693 0.0144 |
| Lion > 2.2 yrs mane length (West [19]) | 442 | 0.1936 0.8064 | 0.6800 1.2663 | 0.0192 0.0338 | 0.7171 0.2829 | 0.2489 -0.2681 | 0.0166 0.0827 |
| Lion > 2.2 yrs mane darkness (West [19]) | 442 | 1.0000 | 1.1008 | 0.0562 | 0.6464 0.3536 | 0.1695 -0.1118 | 0.0217 0.0673 |
| Lion> 5 yrs mane length (West [19]) | 257 | 1.0000 | 1.2977 | 0.0319 | 0.0383 0.9617 | -0.1331 0.2652 | 0.0814 0.0145 |
| Lion > 5 yrs mane darkness (West [19]) | 257 | 1.0000 | 1.2021 | 0.0363 | 0.3205 0.6795 | 0.0484 0.2283 | 0.0351 0.0142 |
| Dung beetle horn length - WA (Moczek [20]) | 644 | 0.3546 0.0837 0.1616 0.1910 0.2091 | 0.5105 2.0758 1.1310 0.6517 3.9032 | 0.0033 0.4042 0.0782 0.0110 0.2139 | 0.4784 0.2111 0.3105 | -0.6237 1.3512 0.1371 | 0.0224 0.0152 0.2152 |
| Dung beetle horn length - NC (Moczek [20]) | 1016 | 0.2301 0.1633 0.2268 0.3799 | 2.6811 0.9594 0.5430 4.0161 | 0.6706 0.0686 0.0097 0.1330 | 0.2423 0.1907 0.2292 0.3378 | 0.1279 1.1523 -0.6075 1.4015 | 0.2082 0.0295 0.0418 0.0064 |
| Earwig forceps length (Tomkins [21]) | 134 | 0.3165 0.2501 0.4333 | 5.9727 7.3120 3.5705 | 0.7099 0.1154 0.0982 | 0.2964 0.2460 0.4576 | 1.8033 1.9901 1.2796 | 0.0144 0.0020 0.0098 |
| Great tit stripe length (Norris [22]) | 63 | 0.5789 0.4211 | -14.1532 17.1432 | 77.5214 60.9468 | NA | NA | NA |
| Fiddler crab fight duration (Hyatt [23]) | 80 | 0.0500 0.4433 0.1848 0.3219 | 482.1489 19.5720 51.5698 125.9917 | 5555.8303 65.1629 33.6134 2189.0050 | 1.0000 | 3.8413 | 1.1077 |
| Fiddler crab fight acts (Hyatt [23]) | 80 | 0.1103 0.2370 0.6526 | 53.5474 26.7320 11.3968 | 555.7647 14.2841 22.7231 | 1.0000 | 2.7213 | 0.5112 |

TABLE II. We fit Gaussian mixture models with 1–5 components of unequal variance to the animal ornamentation data sets and find the number of components that yields the best BIC [6]. We performed this fit on log-transformed data because tissue measurements are often log-normally distributed [24]. Note that we exclude Gaussian mixture models for log-transformed data (NA) if the original data is not a straight-forward measurement of tissue investment.

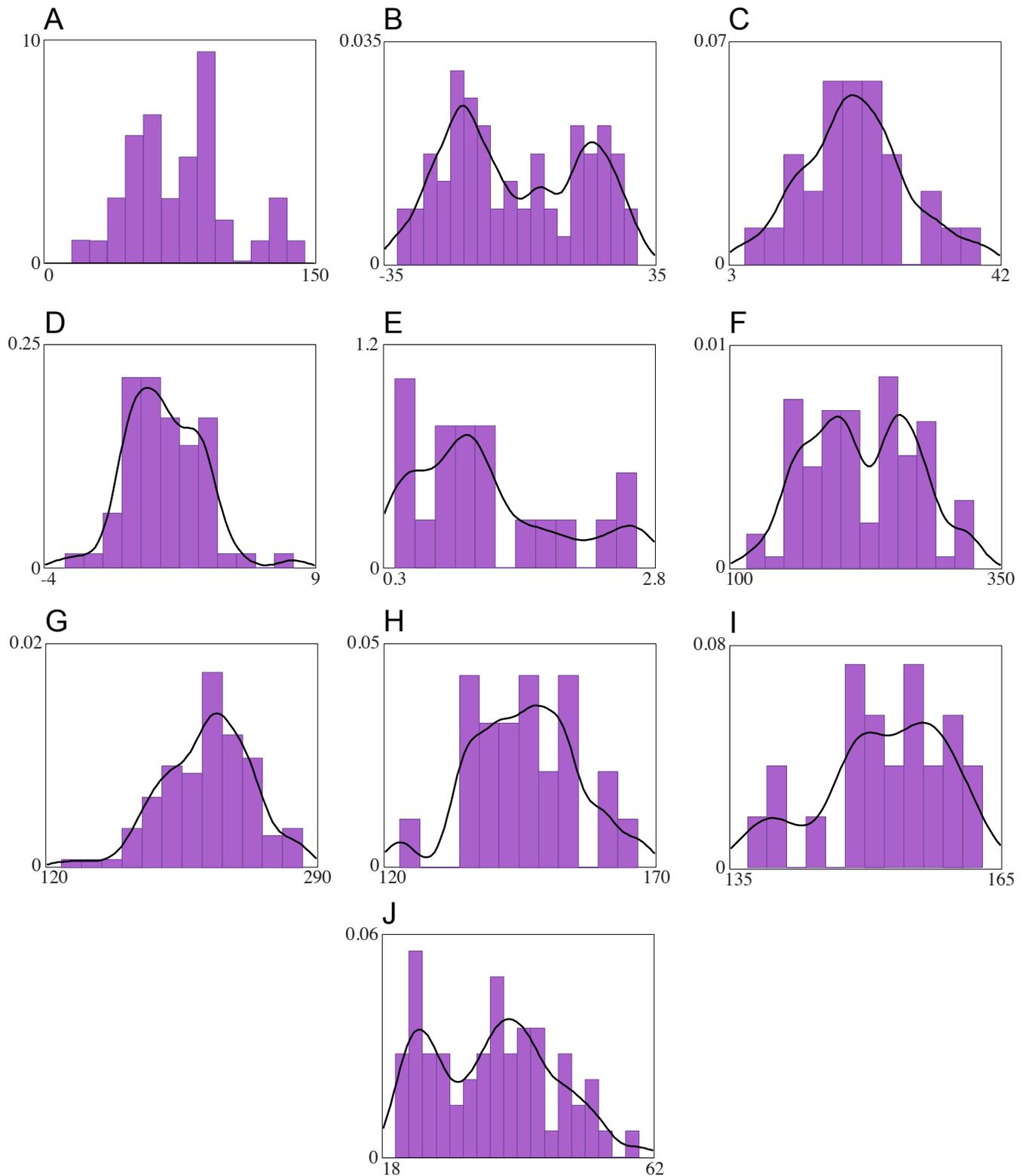

FIG. S5. Additional ornament data sets (birds) **A.** Blackbird song pulse repetition rate [30] (data extracted from histogram, so sample size uncertain) **B.** Great tit stripe size [22] (N=63) **C.** Partridge black ventral area [17] (N=29) **D.** Finch carotenoid coloration [15] (N=68) **E.** Barn owl spottiness [14] (N=20) **F.** Widowbird collar patch size [13] (N=107) **G.** Widowbird tail length [13] (N=107) **H.** Peacock eye spots [10] (N=24) **I.** Peacock eye spots [9] (N=24) **J.** Yellow-breasted chat plumage color [8] (N=62)

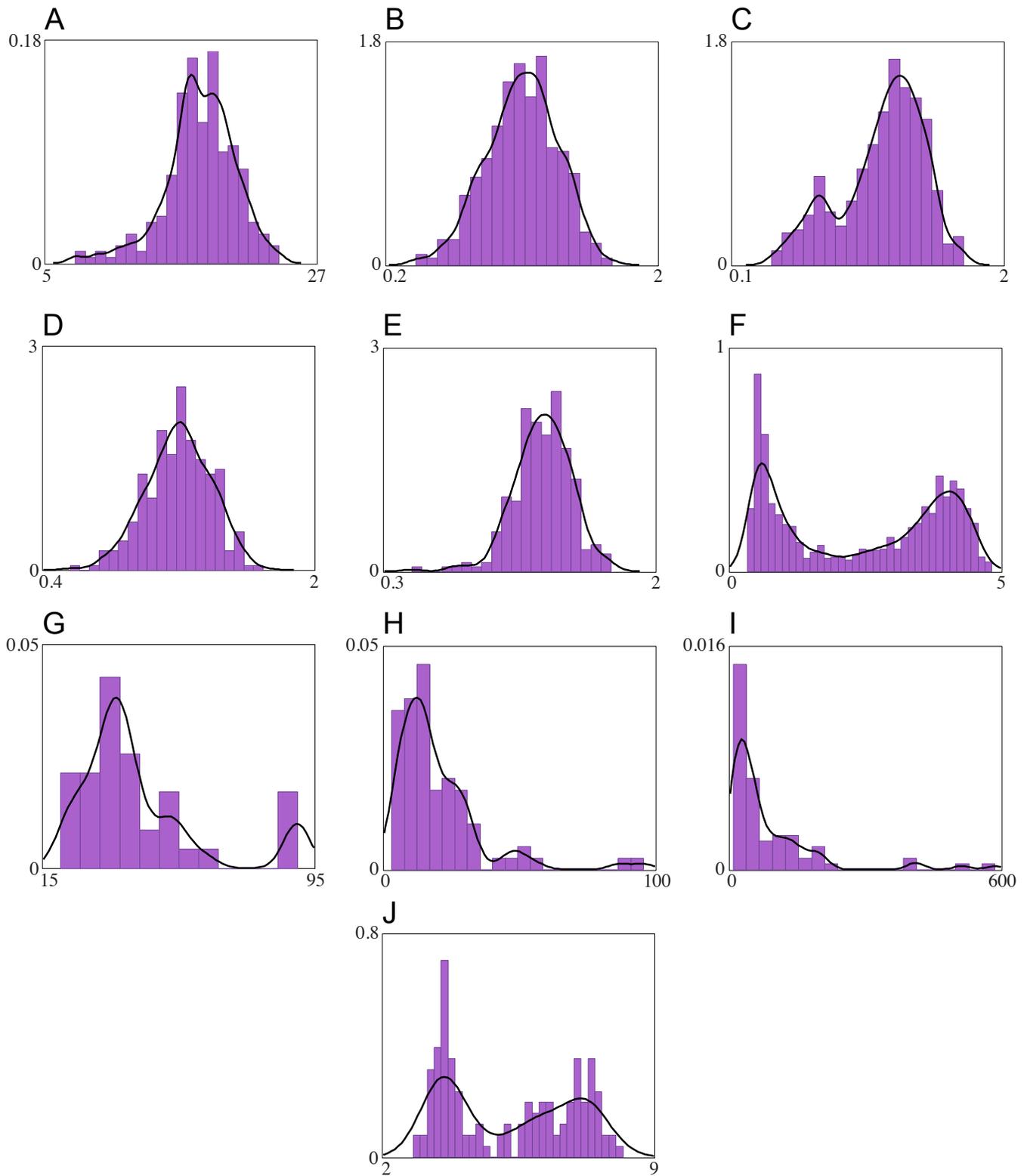

FIG. S6. Additional ornament data sets **A.** Roe deer antler length [18] (N=242) **B.** Mature (> 2.2 yr) lion mane darkness [19] (N=442) **C.** Mature (> 2.2 yr) lion mane length [19] (N=442) **D.** Older (> 5 yr) lion mane darkness [19] (N=257) **E.** Older (> 5 yr) lion mane length [19] (N=257) **F.** Dung beetle horn length (North Carolina) [20] (N=1016) **G.** Stickleback nest compactness [16] (N=38) **H.** Fiddler crab fight acts [23] **I.** Fiddler crab fight duration [23] **J.** Earwig forceps length [21] (N=134)

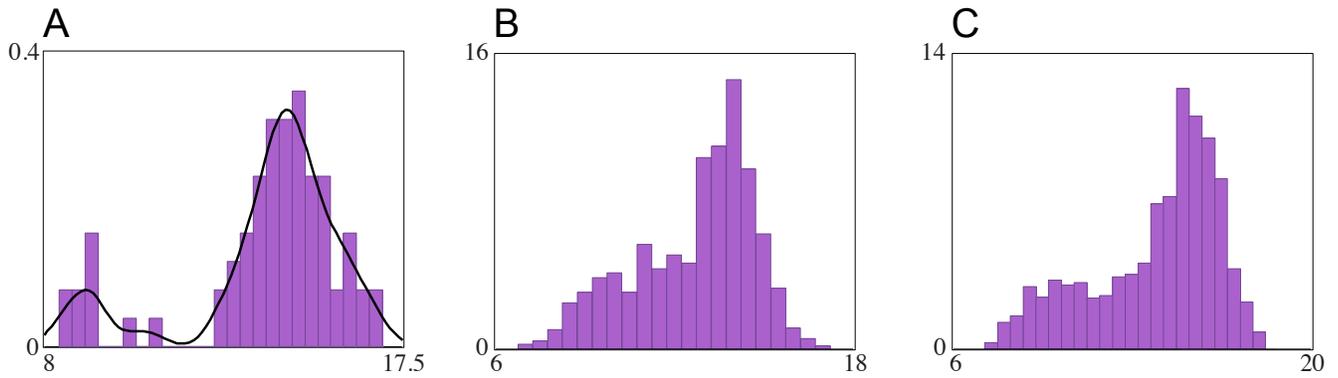

FIG. S7. Bimodal body size data sets **A.** Salmon body size [12] (N=72) **B.** Trout body size (early season) [26] (data extracted from histogram, so sample size uncertain) **C.** Trout body size (late season) [26] (data extracted from histogram, so sample size uncertain)

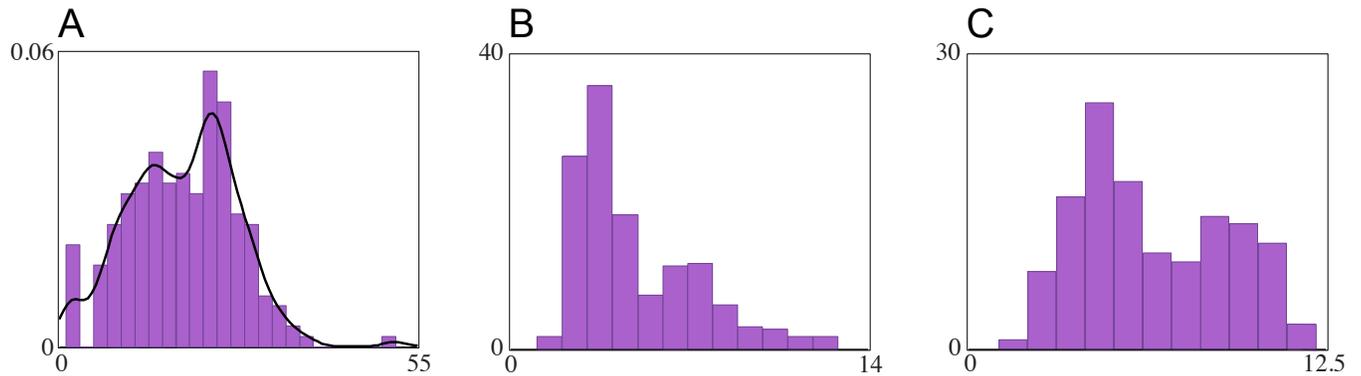

FIG. S8. Bimodal forest data sets **A.** Diameter at breast height for B. platyphylla trees [28] (N=217) **B.** Diameter at breast height for B. ermanii (11-16 yrs old) [29] (data extracted from histogram, so sample size uncertain) **C.** Height of B. ermanii (11-16 yrs old) [29] (data extracted from histogram, so sample size uncertain)



\end{document}